\documentclass[pra,twocolumn,showpacs,aps,floatfix]{revtex4-1}

\usepackage{amsmath}
\usepackage{graphicx}
\usepackage{color}

\graphicspath{{figures/}}

\newcommand{\beq}{\begin{equation}}
\newcommand{\eeq}{\end{equation}}

\begin{document}

\title{Chaotic scattering in the presence of a dense set of overlapping Feshbach resonances}
\author{Krzysztof Jachymski$^1$ 
 and~Paul~S.~Julienne$^2$}
\affiliation{
$^1$ Faculty of Physics, University of Warsaw, Pasteura 5,
02-093 Warsaw, Poland\\
$^2$
Joint Quantum Institute, University of Maryland and National Institute of Standards and Technology, College Park, Maryland 20742, USA}
\date{\today}

\begin{abstract}
Complex quantum systems consisting of large numbers of strongly coupled states exhibit characteristic level repulsion, leading to a non-Poisson spacing distribution which can be described by Random Matrix Theory. Scattering resonances observed in ultracold atomic and molecular systems exhibit similar features as a consequence of their energy level structure. We study how the overlap between Feshbach resonances affects the distribution of resonance spacings. The spectrum of strongly overlapping resonances turns out to be non-Poisson even when the assumptions of Random Matrix Theory are not fulfilled, but the spectrum is also not completely chaotic and tends towards being semi-Poisson.
\end{abstract}

\maketitle
{\it Introduction}.  Magnetic Feshbach resonances are widely used to control atomic interactions in ultracold quantum gases~\cite{Chin2010}. They have been extensively studied theoretically and experimentally for many different combinations of alkali-metal-atoms, for which detailed theoretical models have been constructed to understand the Feshbach spectra that result from tuning the magnetic field. In these cases quantum numbers can readily be assigned to the molecular states that give rise to individual  resonances. Precise experimental control and theoretical understanding combine to allow the production of weakly bound Feshbach molecules~\cite{Kohler2006}. The STIRAP technique~\cite{stirap}, which allows adiabatic transfer of the population to deeply bound states by using properly designed laser pulses, has opened the possibility to study ultracold molecular gases~\cite{Jochim2003,Zwierlein2003,Ni2008,Danzl2008,Molony2014,Takekoshi2014,Park2015}.

Close to an isolated Feshbach resonance caused by a ramping bound state that crosses the collision threshold at energy $E=0$ as magnetic field $B$ is tuned, the $s$-wave scattering length for a near-zero-energy collision that has no inelastic loss is described by the well-known formula~\cite{Moerdijk1995,Timmermans1999}
\beq
a(B)=a_{bg}\left(1-\frac{\Delta}{B-B_{res}}\right),
\eeq
where $\Delta$ is the resonance width and $B_{res}$ is the position of the resonant pole in $a(B)$. However, it is often the case that two or more Feshbach resonances overlap with each other so that a proper description of the scattering properties requires taking all the relevant bound states into account~\cite{Jachymski2013,Wang2014}. This is most pronounced for non-alkali systems such as highly magnetic lanthanide atoms, where anisotropic  interactions introduce strong mixing of many partial waves, which can result  in hundreds of overlapping Feshbach resonances~\cite{Petrov2012,Gonzalez2015}. In experiments performed with erbium~\cite{Frisch2014} and dysprosium~\cite{Baumann2014,Maier2015} atoms, the observed resonance density is of the order of a few resonances per gauss for bosonic atoms with no nuclear spin and exceeds 20 resonances per gauss for the fermion $^{167}$Er, which has additional hyperfine structure. In such cases there is no way to assign quantum numbers to individual resonances or to predict their positions theoretically. Interestingly, the resonances can still be used to produce weakly bound dimers~\cite{Frisch2015}.

Statistical analysis of the Feshbach spectra recorded for erbium~\cite{Frisch2014} showed that the spacings between neighboring resonances exhibit characteristic correlations. While one would expect a Poisson distribution of spacings for uncorrelated energy levels, the experimental results were closer to a Wigner-Dyson distribution (WD)~\cite{Wigner1951,Dyson}, which indicates strong repulsion between the overlapping resonances. This kind of level spacing distribution can be derived by means of random matrix theory (RMT) and has been shown to accurately reproduce resonances in nuclear reactions~\cite{RMT1,RMT2}. Recently, it has been predicted that possible resonance densities in ultracold atom-molecule and molecule-molecule collisions can be so high that they may exhibit similar behavior. References~\cite{Bohn2012,Mayle2013} show how to estimate the density of states and use RMT to construct a statistical scattering model. The Bohigas-Giannoni-Schmit conjecture implies that for systems with such statistical properties the classical dynamics would be chaotic~\cite{BGS}.

On the other hand, the scattering length for an arbitrary set of overlapping Feshbach resonances can also be found analytically, e.g. by using using methods of multichannel quantum defect theory (MQDT)~\cite{Jachymski2013}. In this paper, we apply this analytical model to the case of extremely dense Feshbach spectra. Interestingly, we find that the distribution of the resonance spacings starts to resemble the Wigner-Dyson distribution when the mean resonance width exceeds average spacing between the resonances, even if the RMT assumptions are not fulfilled. 

{\it Scattering length in the presence of many resonances}.
We start our analysis by considering a general scattering problem of two particles with complex internal structure giving rise to a large possible number of scattering resonances. Instead of studying in detail the origin of the resonances as done in~\cite{Petrov2012}, we will simply assume that there are $N\gg 1$ bound states of the composite 2-particle system which can cross with the single entrance channel as $B$ is tuned over a range spanned by these levels 
(it is also possible to consider fields other than a magnetic one to do the tuning).
This channel is characterized by its background scattering length $a_{bg}$. We will assume that the particles interact via a van der Waals potential at long range and express $a_{bg}$ as $r \bar{a}$ with $\bar{a}$ being the mean scattering length of the van der Waals potential~\cite{Gribakin1993}. The characteristic range of interparticle separation at which interchannel couplings are present will be assumed to be much smaller than $\bar{a}$. Under these assumptions, the magnetic field dependence of the scattering length for a set of N overlapping Feshbach resonances can be conveniently described by the following analytical formula~\cite{Jachymski2013}
\beq
\label{asc}
a(B)=a_{bg}\left(1-\sum_{i=1}^N{\frac{\Delta_i}{B-B_i-\delta B_i-\sum_{j\neq i}{\frac{B-B_i}{B-B_j}\delta B_j}}}\right).
\eeq
Here $\Delta_k$ is the local width associated with $k$-th resonance, $B_k$ is the position at which the bare $k$-th bound state crosses the threshold and $\delta B_k$ is the corresponding shift of the resonance position due to interaction with the open channel, for van der Waals interactions given by $\delta B_k=\frac{r(1-r)}{1+(1-r)^2}\Delta_k$. This formula can be derived using the MQDT formalism described in detail in~\cite{Jachymski2013}, where it was able to account for overlapping resonances in several different examples involving alkali-metal atoms. The actual positions of resonances $B_k ^{res}$ can be calculated by finding the zeros of the denominators in~Eq.~\eqref{asc}. In general, the widest and  most closely lying resonances tend to have the strongest impact on the other ones.

In the derivation of this formula, the closed channels are chosen to be uncoupled, with the only nonzero couplings existing between the entrance channel and the closed channels. 
Another assumption of our model is linear variation of the tangent of the bound state phase $\tan\nu_k(E)\propto E-\delta\mu(B-B_k)$ for each closed channel state. As a result of these approximations, our model is applicable only near the energy threshold.
Formula~\eqref{asc} describes the scattering length in terms of ``local'' resonance widths. The $k$-th width $\Delta_k$ is determined only by the coupling with the bound state crossing the threshold at $B_k$. It is also possible to algebraically transform~Eq.~\eqref{asc} to the form
\beq
\label{effasc}
a(B)=a_{bg}\prod_{i=1}^N{\left(1-\frac{\tilde{\Delta}_i}{B-B^{res}_i}\right)}.
\eeq
Here the resonance widths are a function of \textit{all} interchannel couplings instead of local parameters. Regardless of the choice, the resonance positions are the same and it is the distribution of $B_{res}$ and the spacings between them that is of our current interest. 

{\it Dense sets of resonances}.
Our MQDT formalism successfully reproduced the complex structure of $a(B)$ when several overlapping resonances occur in the case of alkali metal species~\cite{Jachymski2013}.  It can also be used to set up a test system for a dense set of resonances in which we can examine the change in level statistics that occurs as the coupling strength increases, with different assumptions about the distribution of the interactions $\Delta_k$ or of the ``bare'' crossing positions $B_k$ before coupling to the entrance channel is turned on.  Let us emphasize that we are not intending to produce realistic models for atomic species like Er or Dy, since our model of the potentials and interactions is too simple to capture the physics of such complex atoms.  Rather, we are interested in understanding some of the conditions that can give rise to Poisson or WD distributions when we have a set of interacting tunable Feshbach resonances in cold atomic collisions.

The conventional way of dealing with systems with large number of coupled channels with no good quantum numbers (apart from time-reversal symmetry) is to use RMT. In this approach all the couplings are assumed to be random numbers with statistical properties governed by Gaussian Orthogonal Ensemble. As a result, the energy level spacings as well as resonance spacings follow the Wigner-Dyson distribution, defined as
\beq
p_{WD}(s)=\frac{\pi s}{2} \mathrm{exp}\left(-\frac{\pi s^2}{4}\right),
\eeq
where $p_{WD}(s)$ gives the probability density and $s$ is the distance between adjacent levels in units of mean resonance spacing $d=1/\rho$ with $\rho=N/B_{max}$, where we assume $B$ to vary from 0 to $B_{max}$ for convenience. While normally the Hamiltonian of the system is fixed and the distribution of energy levels is characterized, in studying Feshbach resonance spacings we fix $E=0$ and look at the resonance distribution as the Hamiltonian is varied by changing $B$.

Within our treatment we start from Eq.~\eqref{asc} and make different statistical assumptions about the bare resonance positions and coupling strengths. In the simplest scenario one can assume all couplings to be equal and all bare resonance positions to be completely uncorrelated. The distribution of spacings between adjacent bare levels is then a Poisson one that can be described by
\beq
p_{P}(s)=e^{-s}
\eeq
This clearly violates the assumptions of RMT. The correlations between observed resonance positions will come solely from the terms $\frac{B-B_i}{B-B_j}\delta B_j$ in Eq.~\eqref{asc} that describe mutual interaction between the resonances. For very weak couplings, resulting in mean $\Delta$ (and consequently $\delta B$) much smaller than average distance between the resonances, the distribution of spacings should intuitively stay a Poisson one. However, for large widths the resonances will start to overlap significantly and the statistics will be modified.
For a given set of resonance parameters $\{r,\Delta_i,B_i\}$ we obtain the statistics of resonance spacings numerically and fit it to the Brody distribution, defined as~\cite{Brody}
\beq
p_{B}(s)=\alpha(\eta+1)s^\eta \mathrm{exp}\left(-\alpha s^{(\eta+1)}\right),
\eeq
where $\eta$ is a real number $0\leq \eta \leq 1$, $\alpha=\Gamma\left(\frac{\eta+2}{\eta+1}\right)^{\eta+1}$ and $\Gamma$ is the Gamma function. This is a heuristic distribution that in the limit $\eta\to 1$ this function approaches WD distribution, while at $\eta\to 0$ it recovers Poisson statistics. Here we treat $\eta$ as a fitting parameter. Then we average $\eta$ over many numerical realizations corresponding to different sets of $B_i$.

\begin{figure}
\centering
\includegraphics[width=0.5\textwidth]{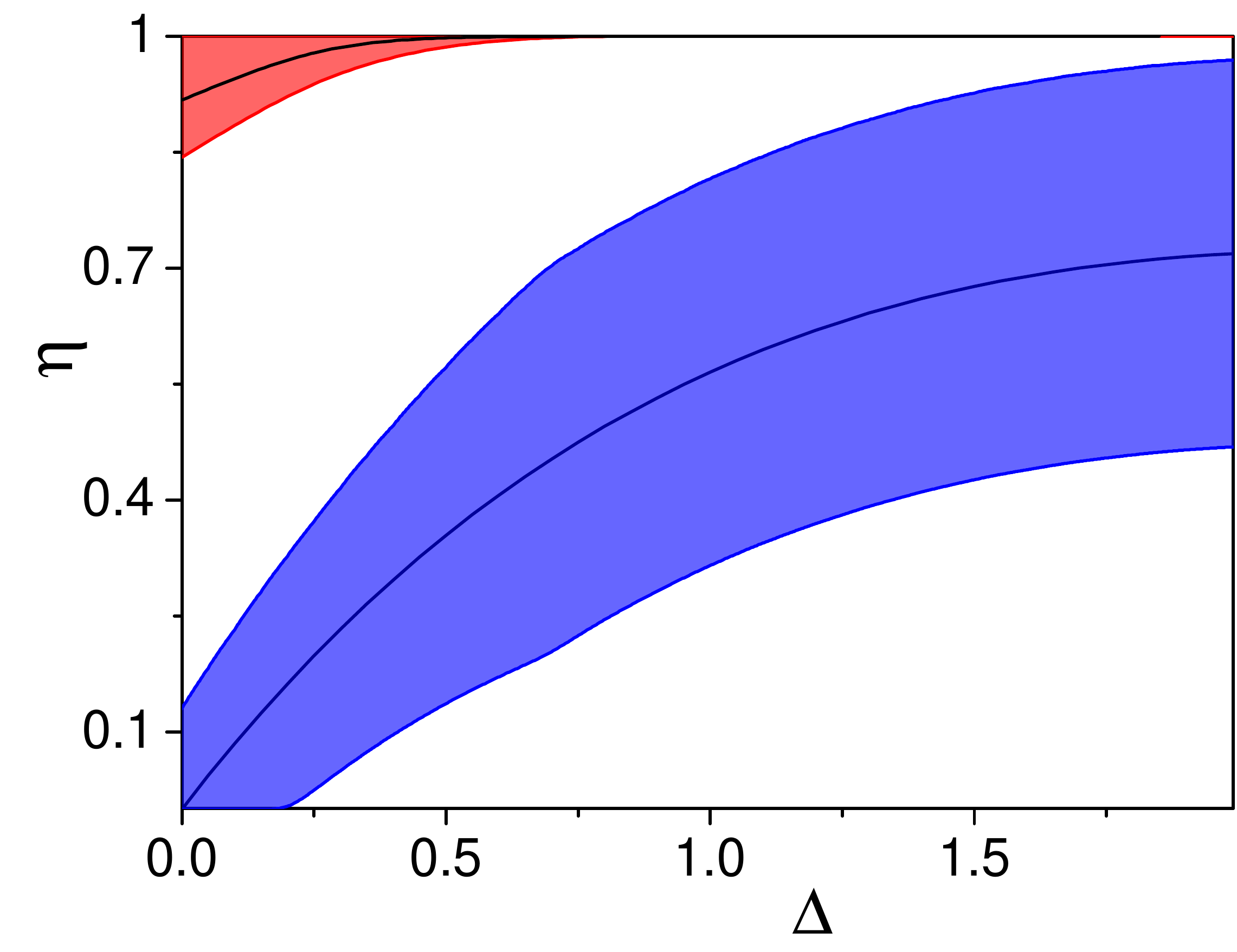}
\caption{\label{figbrody}(color online) Brody parameter fitted to a set of resonances with width $\Delta$ in units of mean spacing,  averaged over 25 realizations, along with the standard deviation for a single realization (shaded regions). Blue: Poisson initial distribution of resonance positions, red: Wigner-Dyson initial distribution of resonance positions.}
\end{figure}

{\it Results and discussion}. Figure~\ref{figbrody} summarizes our findings for an example with 50 resonances of equal widths $\Delta$, showing the dependence of Brody parameter $\eta$ on the width of the resonances. For uncorrelated bare resonance positions (blue line), we observe that at small widths (less than the mean spacing $d$), the spacing distribution is essentially a Poisson one, but in the limit of large widths $\Delta\gg d$ it tends towards a WD distribution with the value of $\eta$ being around $0.7$. It is worth noting that this is close to the value measured in experiments with ultracold Er and Dy~\cite{Frisch2014,DyErFB}, although our model assumptions are not designed to represent these species. We checked that this feature is independent of  $a_{bg}$. This approach of $\eta$ towards values of around 0.7 still holds if we assume a distribution of initial resonance widths, although when one starts from a set containing many very narrow resonances which do not influence each other strongly, the effect will be naturally weaker and will require much larger average $\Delta$. On the other hand, if the bare resonance positions come from a WD distribution, the additional effect of the coupling between resonances due to the width $\Delta$ in~\eqref{asc} changes the statistical properties only very slightly, as shown by the red line in Figure~\ref{figbrody}. 

\begin{figure}
\centering
\includegraphics[width=0.5\textwidth]{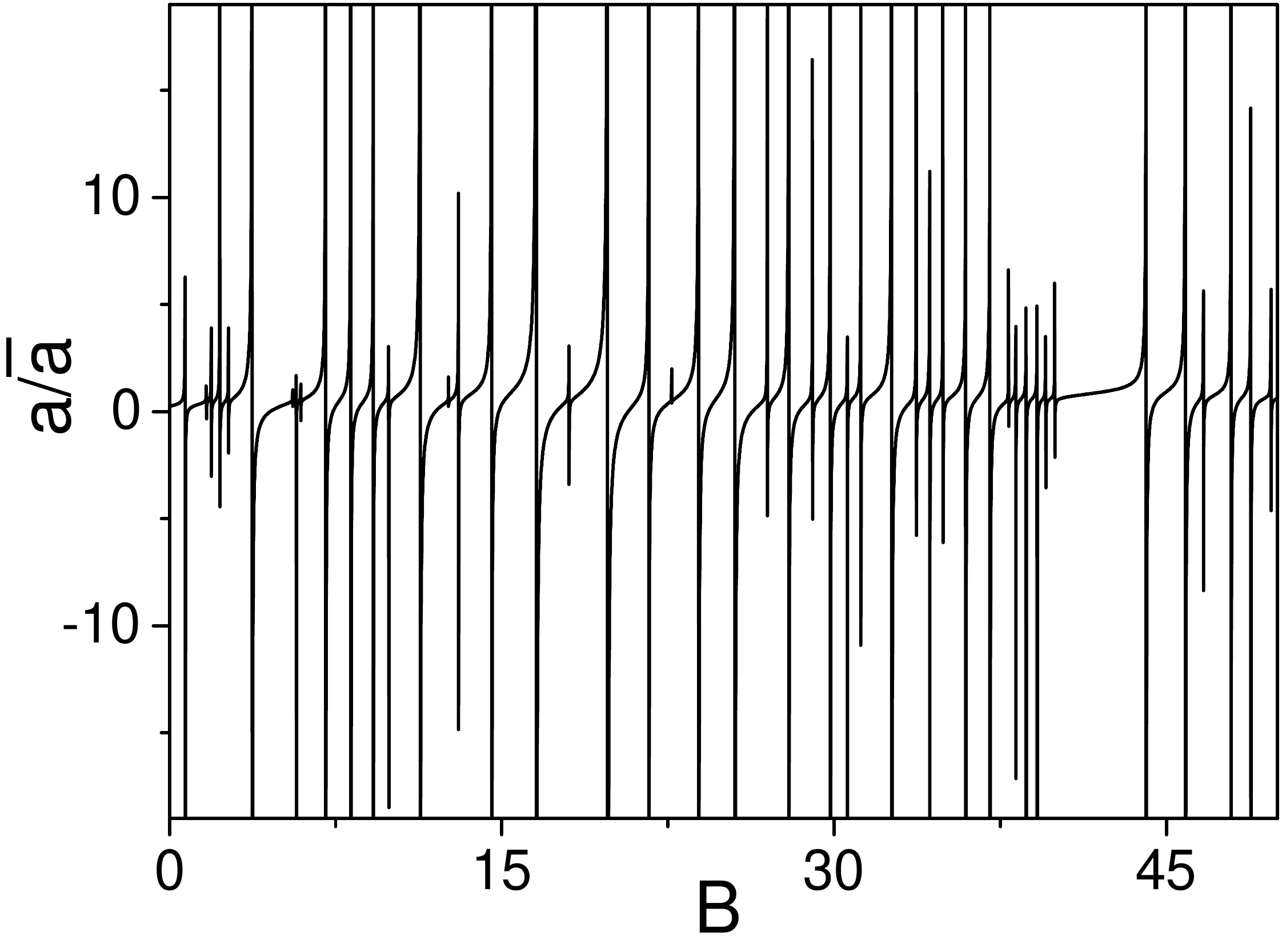}
\caption{\label{figasc}Scattering length versus $B$ in units of the mean spacing $\rho$ for an example with 50 resonances having equal widths of twice the mean spacing. Not every resonance is visible as a divergence of $a$ due to finite resolution of the calculated set of points.}
\end{figure}

\begin{figure}
\centering
\includegraphics[width=0.5\textwidth]{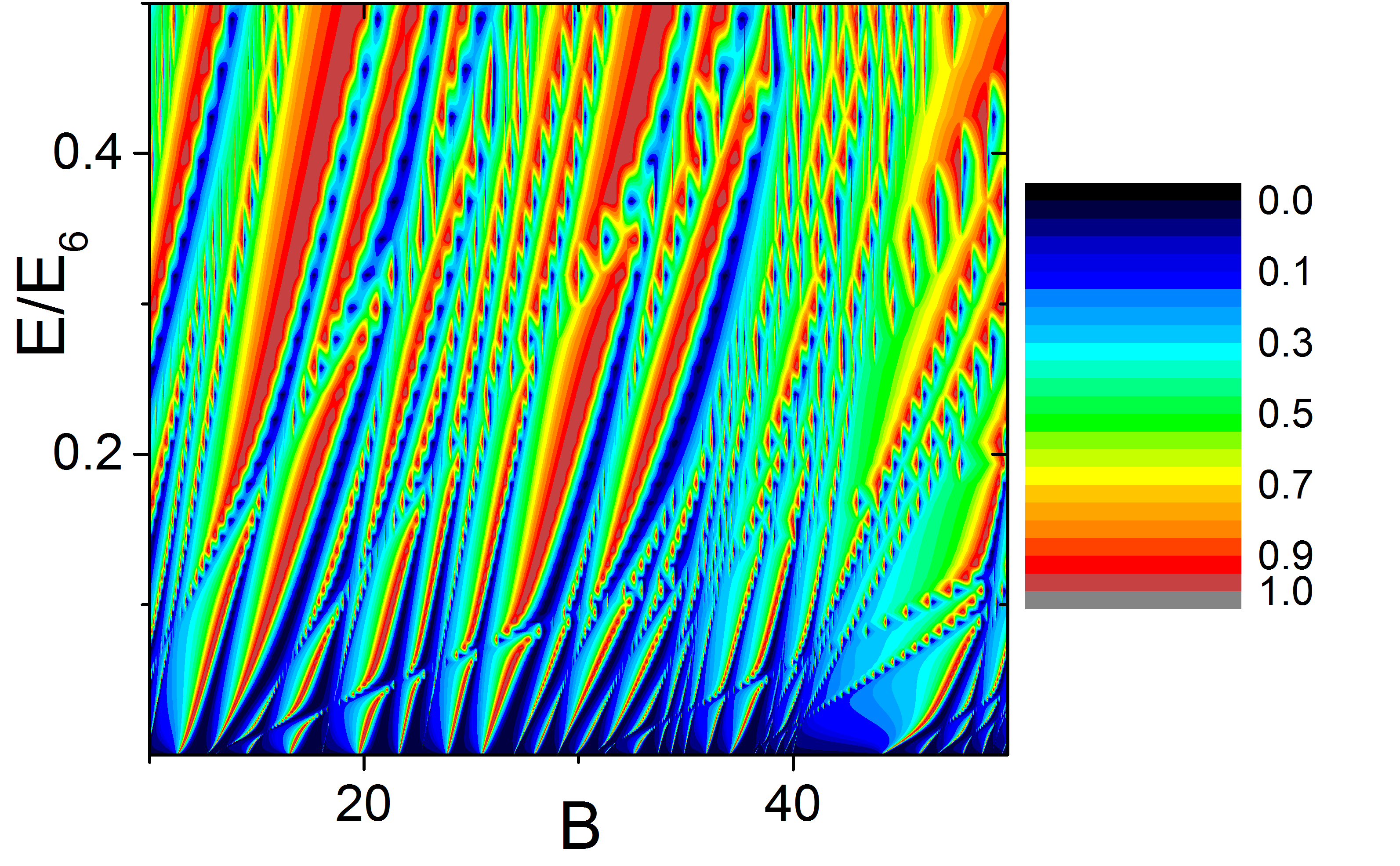}
\caption{\label{figsin}(color online) $\sin ^2 \delta(E)$ for the same set of resonances as on Fig.~\ref{figasc}, assuming random magnetic moment differences ranging from $0$ to $0.5\,E_6$ per magnetic field unit.}
\end{figure}

It is interesting to study the scattering length $a(B)$ as well as finite energy scattering properties for an example case. Figure~\ref{figasc} shows the scattering length for the case of 50 strongly overlapping resonances with an initial Poisson distribution and equal widths. Note that the widths of the interacting resonances are not the same, but show considerable variation due to interactions among the resonances. When the initial resonance width exceeds strongly the mean resonance spacing $d$, the average observed resonance width should approach the value of $d$. As a result, many narrow resonances can be seen on Figure~\ref{figasc}. For weakly overlapping resonances the widths are not modified so strongly. In contrast to typical situations in alkali gases, it is hard to extract a unique value of $a_{\rm bg}$ from the figure. In fact, we find that the mean scattering length calculated over a range of magnetic field will take a value of the order of $\bar{a}$, the mean scattering length of the van der Waals potential. Interestingly, two recent experiments with Dy atoms~\cite{Maier2015,Lev2015} both reported finding background scattering lengths having the order of magnitude of this mean scattering length.

The MQDT theory of Ref.~\cite{Jachymski2013} analytically continues the bound state spectrum and threshold properties into the finite $E$ scattering continuum, where resonances can cross one another and the resulting scattering phase shift can exhibit a complex pattern. An additional parameter $\delta\mu$ for each resonance, describing the magnetic moment difference between the open and closed channel states, is needed for this extension. Figure~\ref{figsin} shows the value of the sine square of the scattering phase shift, $\sin ^2 \delta(E)$, at different energies and magnetic fields for the same set of resonances as on Figure~\ref{figasc}, assuming $\delta\mu$ parameter to be a random number between 0 and 0.5 $E_6$ per magnetic field unit. The Figure shows that at finite energies the phase shift varies rapidly with both energy and magnetic field as narrow continuum levels cross one another. Such behavior would have notable consequences, making the scattering properties of a thermal gas (characterized by wide distribution of collision energies) much harder to control.
 
\begin{figure}
\centering
\includegraphics[width=0.5\textwidth]{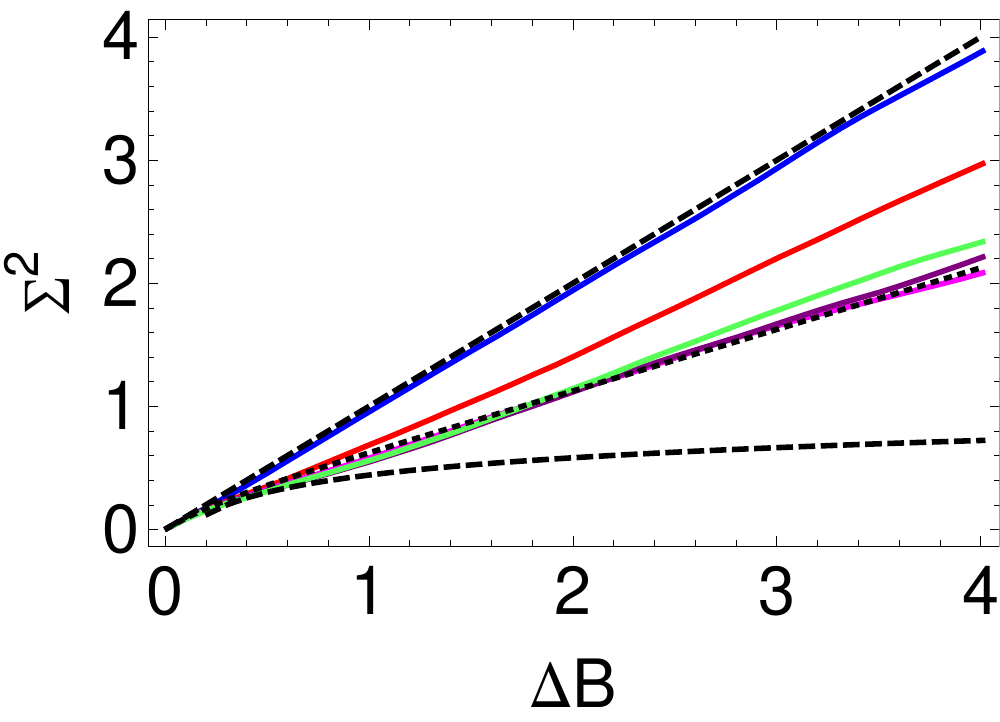}
\caption{\label{figsigma}(color online) Number variance $\Sigma^2$ of the resonance positions for five different resonance widths, $\Delta= 0.2$ (blue), $0.5$ (red), $1$ (magenta), $1.5$ (purple) and $2.5$ (green) in units of mean spacing $d$, averaged over 25 realizations. The dashed lines show the limiting analytical results for WD (lower line) and Poisson (upper line) statistics. The dotted line depicts the semi-Poisson distribution.}
\end{figure}

In addition to the level spacing distribution, the fluctuations in the number of resonances among different magnetic field intervals are commonly studied for complex systems. They can be measured by the number variance $\Sigma^2(\Delta B) = \overline{N^2}(\Delta B) -\overline{N(\Delta B)} ^2$, where $N(\Delta B)$ is the number of resonances found in the range from $B_0$ to $B_0+\Delta B$ and the averaging is performed over $B_0$. Figure~\ref{figsigma} shows the results obtained for our model for different resonance widths. The spectra are far more rigid than a Poisson distribution would predict but are still quite far from the predictions of WD statistics. In fact, we found that the semi-Poisson distribution, $p_{SP}(s)=4se^{-2s}$, for which the number variance $\Sigma^2(\Delta B)=\Delta B/2+(1-e^{-4\Delta B})/8$~\cite{Garcia2005}, characterizes the statistical properties of our model much better than the Poisson or Wigner-Dyson one. The semi-Poisson distribution describes the statistics directly at the metal-insulator phase transition in condensed matter systems~\cite{Shklovskii1993,Garcia2005}. Fitting a sample of random numbers with semi-Poisson distribution to the Brody distribution gives the $\eta$ parameter of around 0.6, in qualitative agreement with the behavior we observed on Fig.~\ref{figbrody}.

{\it Conclusions}.
We have extended an analytic MQDT model that successfully explains sparse sets of overlapping resonances of alkali-metal species to examine the statistics of level spacings of dense sets of many overlapping resonances.  While such a model is not designed to capture the physics of complex atomic species like Er or Dy, it demonstrates the onset of a level spacing distribution similar to the chaotic one.  If a non-interacting set of resonances described by a Poisson distribution of spacings is coupled to an ultracold $s$-wave entrance channel with a set of widths that are small compared to the mean level spacing, then the interacting set maintains a near-Poisson distribution.  However, when the widths begin to exceed the mean level spacing, the level spacing distribution changes, with a Brody parameter $\eta$ that has a value of around 0.7 and is similar to that observed for species like Er or Dy. This happens in spite of the fact that the MQDT model makes no assumptions about random couplings as in Random Matrix Theory. The differences between the predictions of our model and the Wigner-Dyson statistics are best seen when analyzing the number variance, which for our case is well described by a semi-Poisson distribution, even for very strongly overlapping resonances.  Thus, the current observational and theoretical evidence on the level spacing and its number variance~\cite{DyErFB} combined with our result suggest that a semi-Poisson distribution may provide a better approximation to these quantities than a Wigner-Dyson one.  This is a question that could be explored in future research.

Our results show that correlations between resonance positions in ultracold collisions can have a different source than a chaotic distribution of bound levels. The effect of overlapping of resonances itself introduces level repulsion which will affect the analysis of experimental data. Another particular experimental and theoretical challenge lies in measuring and interpreting the resonance widths. Recent analysis of Er data~\cite{Mur2015}, based on the width of three-body loss features, showed discrepancies with RMT predictions. However, the three-body recombination process in the vicinity of narrow overlapping resonances is very difficult to describe precisely~\cite{Wang2014} and it is not obvious what information can be extracted from this about the resonance widths.

The MQDT model presented here can be extended to take into account more detailed assumptions about the interaction potentials and provide an analysis of realistic species complementary to the one provided by ab initio calculations. For example, ref.~\cite{Maier2015} shows that special non-chaotic eigenstates can emerge from the dense set of chaotic states and play a dominant role in determining the near-threshold bound state and scattering properties. Our MQDT method readily explains the effect of such unique non-chaotic states on the bound  state and scattering properties within an overall dense set of overlapping chaotic resonances. Overall, a variety of theoretical approaches is needed to fully understand the role of quantum chaos in ultracold collisions



This work was supported by the Foundation for Polish Science START programme, International PhD Projects Programme co-financed by the EU European Regional Development Fund and National Center for Science Grant No. DEC-2013/09/N/ST2/02188.

\bibliography{Allrefs}
\end{document}